\documentclass[11pt,twoside]{article}

\usepackage{asp2006}
\usepackage{epsf}

\begin{document}

\title{Simulations of Line Profile Structure in Shell Galaxies}   

\author{
Lucie J\'ilkov\'a,$^1$
Bruno Jungwiert,$^2$
Miroslav K\v{r}\'i\v{z}ek,$^{2,3}$
Ivana Ebrov\'a,$^{2,3}$
Ivana Stoklasov\'a,$^2$
Tereza Bart\'akov\'a,$^1$
and Kate\v{r}ina Barto\v{s}kov\'a$^1$
}

\affil{$^1$Dept. of Theoretical Physics and Astrophysics, Faculty of Science, Masaryk University, Kotl\'a\v{r}sk\'a 2, CZ-611\,37 Brno, Czech Republic}
\affil{$^2$Astronomical Institute, Academy of Sciences of the Czech Republic, Bo\v{c}n\'{i} II 1401/1a, CZ-141 31 Prague, Czech Republic}
\affil{$^3$Faculty of Mathematics and Physics, Charles University in Prague, Ke~Karlovu 3, CZ-121 16 Prague, Czech Republic}

\begin{abstract} 
In the context of exploring mass distributions of dark matter haloes in giant ellipticals, we extend the analysis carried out by\,\citet{merrifield98} for stellar line profiles of shells created in nearly radial mergers of galaxies. We show that line-of-sight velocity distributions are more complex than previously predicted. We simulate shell formation and analyze the detectability of spectroscopic signatures of shells after convolution with spectral PSFs. 
\end{abstract}

\section{Introduction}
Stellar shells observed in some elliptical galaxies are thought to be by-products of galaxy mergers, predominantly of those involving a giant elliptical with a~much smaller galaxy, e.g.,~a spiral or a dwarf elliptical. The most regular shell systems, Type\,I shell galaxies, are believed to result from a nearly radial merger. Stars of the secondary galaxy oscillate in the potential of the primary and cumulate near the turning points of their orbits. This can be observed as shell-like enhancements of surface brightness if observed along a line-of-sight nearly perpendicular to the merger axis. While the mechanism of shell formation was explained nearly three decades ago \citep{quinn84,dupraz86,hernquist88}, recent discoveries -- e.g.,~a regular shell system in a quasar host galaxy \citep{canalizo07,bennert08}, shells found in M31 \citep{fardal07,fardal08} \mbox{--} bring fresh wind into this  field.

On top of the new data, the shells attract interest due to the (so far theore\-tical) possibility of using them to probe the dark matter distribution of the host galaxy. While \citet{dupraz87} showed that using shell spa\-cing from photometry to constrain the matter distribution is hopeless due to the effects of dynamical friction, \citet[][hereafter MK98]{merrifield98} proposed a way to use spectroscopy to reach the same goal via studying profiles of stellar absorption lines. Here, we extend their analysis beyond monoenergetic shells and show that line profiles from more realistic shells are more complex.

\section{Monoenergetic Shells: Double-peaked LOSVDs}

\begin{figure}
\plottwo{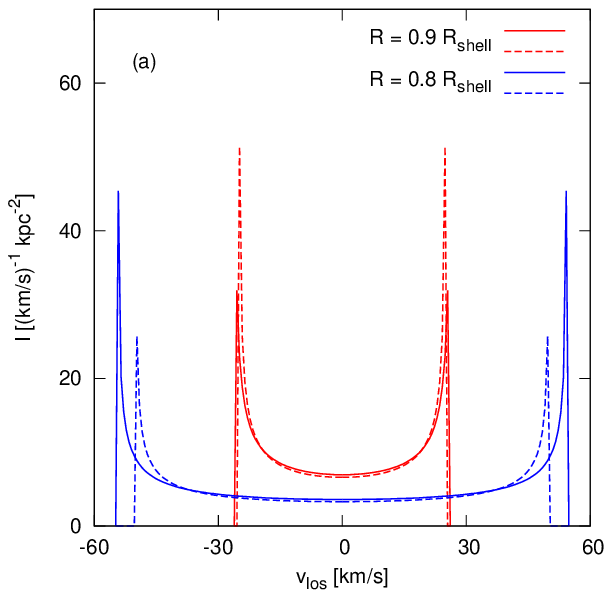}{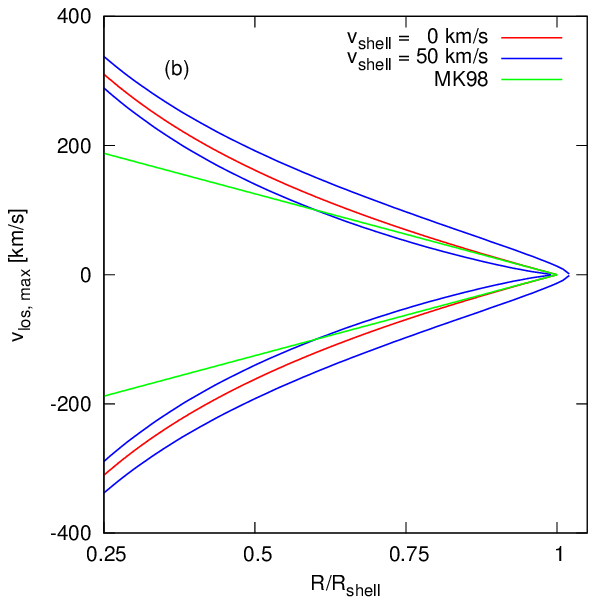}
\caption{
(a) LOSVDs for a monoenergetic shell ($R_{\mathrm{shell}}$\,$=$\,20\,kpc) in the Plummer potential (mass of 3.2$\cdot$10${^{11}}$\,M$_{\odot}$, scaling length of 5\,kpc) at projected radii of 0.9\,$R_{\mathrm{shell}}$ and 0.8\,$R_{\mathrm{shell}}$ (red and blue solid lines). The dashed lines show MK98 approximation.
(b)~$v_{\mathrm{los,\,max}}$ for the monoenergetic, i.e.,~stationary, shell (red line), and the uniformly expanding shell (blue lines) in the Plummer potential as in Fig.\,\ref{mono}a. At the given instant, $R_{\mathrm{shell}}$ is the same for both cases. The green line shows the MK98 approximation.
}\label{mono}
\end{figure}

MK98 studied the kinematics of a monoenergetic shell -- sphe\-ri\-cal system of stars oscillating on radial orbits of the same amplitudes in a spherical potential. The amplitude of oscillations corresponds to the shell-edge radius $R_{\mathrm{shell}}$. They derived an analytic approximation for the line-of-sight ve\-lo\-ci\-ty distribution (LOSVD) in the vicinity of the shell-edge, predicting a~double-peaked profile (Fig.\,\ref{mono}a). The separation of the peaks is related to the gravitational potential of the primary galaxy. For a general gravitational potential and a general projected radius, the LOSVD has no analytical form. We computed the LOSVD numerically as a~ge\-ne\-ra\-li\-za\-tion of the MK98 approach for various gravitational potentials (Plummer, isochrone, de~Vaucouleurs). An example for the Plummer sphere, and two different projected radii, is presented in Fig.\,\ref{mono}a. 

\section{Traveling Shells: Splitting of LOSVD Peaks}

Real shells are not stationary features: the infalling galaxy stars have a con\-ti\-nuous energy distribution, and therefore the shell edge is successively formed by stars of different energies, which appears as the shell edge traveling outwards from the primary-galaxy center. We studied numerically line-of-sight velocity $v_{\mathrm{los}}$ of particles in a uniformly expanding spherical shell. The LOSVD contains signatures of stars returning from a radius where the shell-edge was at some past time, and those traveling to a~position which the shell-edge will reach at a future time. This leads to splitting of both $v_{\mathrm{los}}$
maxima ($v_{\mathrm{los,\,max}}$) at a given projected radius (Fig.\,\ref{mono}b). The stars traveling to their apocenters have higher energies and higher $v_{\mathrm{los,\,max}}$ than the falling stars.

\section{LOSVDs from N-body Simulations}

\begin{figure}[t]
\plotone{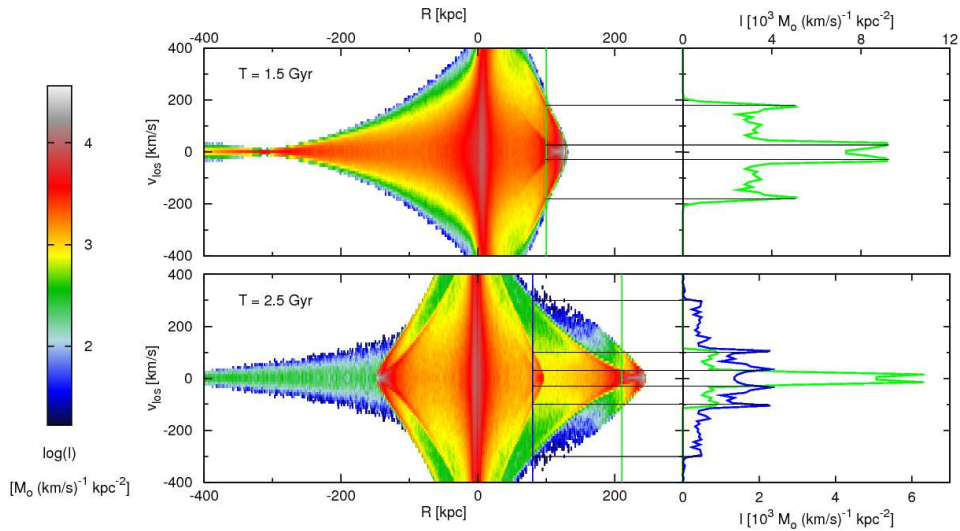}
\caption{Position-velocity maps of particles originally belonging to the secondary galaxy, at two different times. Surface density of particles from the vicinity of merger axis per $v_{\mathrm{los}}$ is mapped. The ``wedges'' correspond to stars traveling to the shell-edge future position and returning from the past one –- notice the splitting similar to Fig.\,\ref{mono}b. Panels on the right represent the LOSVDs (cuts parallel to the velocity axis). The green line corresponds to the outermost (oldest) shell at both times, the blue line to the cut of a younger shell. All cuts are made at the same relative radius 0.8\,$R_{\mathrm{shell}}$.}
\label{mapvel}
\end{figure}

\begin{figure}[t]
\plottwo{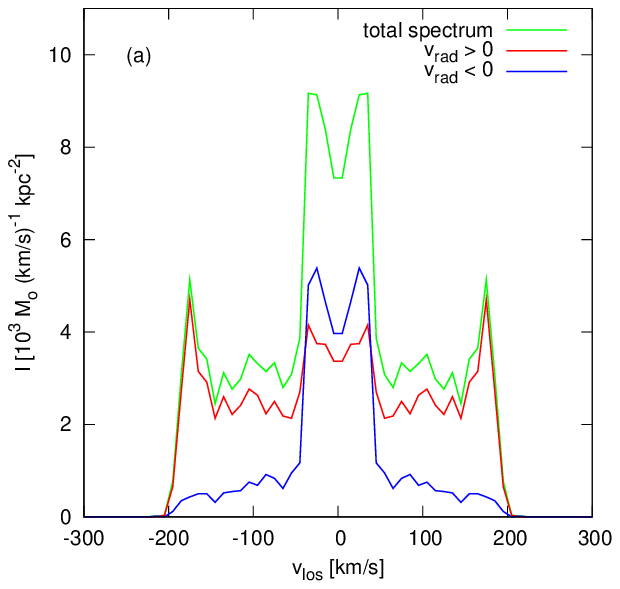}{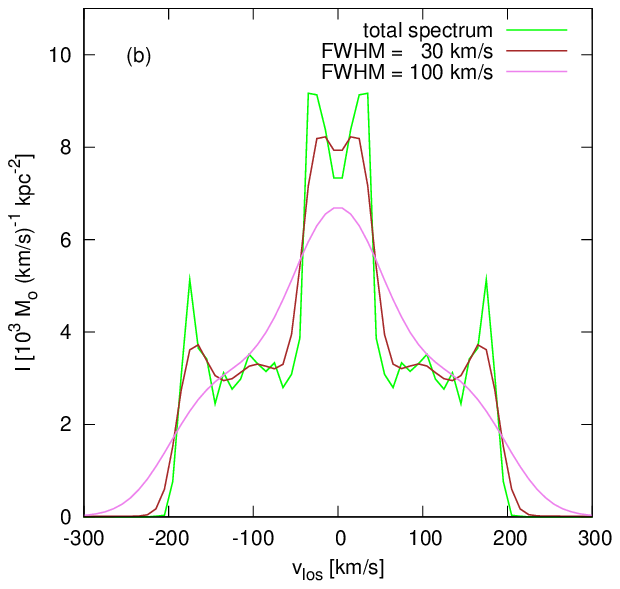}
\caption{(a) Decomposition of the LOSVD (green) to the con\-tri\-bu\-tions produced by stars moving radially outward (red) and inward (blue) with respect to the primary-galaxy center. The same LOSVD of the outermoust shell as in top panel of Fig.\,\ref{mapvel} was used. 
(b)~A~prediction of observed line profiles: green line shows the simulated LOSVDs (same as in Fig.\,\ref{profiles}a), brown and pink lines show convolutions with different Gaussians representing the instrumental dispersion of  FWHM 30 and 100\,km$/$s.}
\label{profiles}
\end{figure}

To study LOSVDs in more detail, we carried out a restricted N-body simulation of shells resulting from a radial merger of a giant elliptical galaxy with a dwarf elliptical (Figs.\,\ref{mapvel} and \ref{profiles}). The primary was represented by a~two-component potential: stars and the dark matter halo. For simplicity the Plummer profile was assumed for both components (with masses of 2$\cdot$10${^{11}}$\,M$_{\odot}$ and 1.2$\cdot$10${^{13}}$\,M$_{\odot}$, scaling lengths of 5\,kpc and 100\,kpc for stars and the dark matter halo, respectively). The dwarf elliptical was simulated as a single Plummer sphere (mass of 2$\cdot$10${^{10}}$\,M$_{\odot}$, scaling length of 2\,kpc).

Right panels in Fig.\,\ref{mapvel} show the LOSVDs for different times of the simulation.  For the outermost shell, we can see a narrowing of the line profile with time, i.e.,~with increasing shell-edge radius, due to the spatial change of the primary's gra\-vi\-ta\-tio\-nal potential (see MK98). The bottom right panel in Fig.\,\ref{mapvel} also shows the inner shell profile, which is more complicated, as it also contains signatures of particles belonging to the outer shell. In Fig.\,\ref{profiles}a, the LOSVD from the top panel of Fig.\,\ref{mapvel} is decomposed according to the sense of particle's motion.

\section{Conclusions}

Theoretical studies of line profiles are needed and timely since getting high S$/$N and high spectral resolution spectra from faint external parts of ellipticals becomes within the reach of current large telescopes. We predict the shape of spectral lines for Type\,I shell galaxies: quadruple-peaked profile. The con\-nec\-tion of this shape with the shell galaxy's gravitational potential is not as straightforward as previously predicted. We also show that relatively high spectral resolution is necessary for observing the line profiles (Fig.\,\ref{profiles}b). To make our study still more realistic, better models for galaxy potentials, and the dynamical friction need to be applied (see \citeauthor{ebrova09}, these proceedings).

\acknowledgements This project is supported by the Institutional Research Plan No.~AV0Z10030501 of the Academy of Sciences of the Czech Republic, by the Doctoral Grant No.~205/08/H005 of the Czech Science Foundation and by the grant LC06014 (Center for Theoretical Astrophysics) of the Czech Ministry of Education.


\begin{thebibliography}{}

\bibitem[Bennert et al.(2008)]{bennert08}
Bennert, N., Canalizo, G., Jungwiert, B., et al. 2008, ApJ, 677, 846

\bibitem[Canalizo et al.(2007)]{canalizo07}
Canalizo, G., Bennert, N., Jungwiert, B., et al. 2007, ApJ, 669, 801

\bibitem[Dupraz \& Combes(1986)]{dupraz86}
Dupraz, C., \& Combes, F. 1986, A\&A, 166, 53

\bibitem[Dupraz \& Combes(1987)]{dupraz87}
Dupraz, C., \& Combes, F. 1987, A\&A, 185, 1

\bibitem[Ebrov\'a et al.(2009)]{ebrova09}
Ebrov\'a, I., Jungwiert, B., Canalizo, G., \& Bennert, N. 2009, these proceedings

\bibitem[Fardal et al.(2007)]{fardal07}
Fardal, M. A., Guhathakurta, P., Babul, A., \& McConnachie, A. W. 2007, MNRAS, 380, 15

\bibitem[Fardal et al.(2008)]{fardal08}
Fardal, M. A., Babul, A., Guhathakurta, P., et al. 2008, ApJ, 682, L33

\bibitem[Hernquist \& Quinn(1988)]{hernquist88}
Hernquist, L., \& Quinn, P.\,J. 1988, ApJ, 331, 682

\bibitem[Merrifield \& Kuijken(1998)]{merrifield98}
Merrifield, M.\,R., \& Kuijken, K. 1998, MNRAS, 297, 1292

\bibitem[Quinn(1984)]{quinn84}
Quinn, P.\,J. 1984, ApJ, 279, 596

\end{thebibliography}
\end{document}